\documentclass[aps,prl,twocolumn,showpacs,floats,floatfix,letterpaper,nofootinbib,superscriptaddress,]{revtex4}

\usepackage{epsfig}

\begin{document}

%%%%%%%%%%%%%%%%%%%%%%%%%%%%%%%%%%%%%%%%
%%%%%%%%%%%%%%%%%%%%%%%%%%%%%%%%%%%%%%%%

\title{Harrison-Zel'dovich primordial spectrum is consistent with observations}

\author{Stefania~Pandolfi}
\affiliation{ICRA and INFN, Universita' di Roma 
	``La Sapienza'', Ple.\ Aldo Moro 2, 00185, Rome, Italy}
\author{Asantha~Cooray}
\affiliation{Center for Cosmology, Department of Physics \& Astronomy, 
	University of California, Irvine, California 92697, USA}
\author{Elena~Giusarma}
\affiliation{IFIC, Universidad de Valencia-CSIC, 46071, Valencia, Spain}
\author{Edward~W.~Kolb}
\affiliation{Department of Astronomy \& Astrophysics, Enrico Fermi 
        Institute, and Kavli Institute for Cosmological Physics,
       	University of Chicago, Chicago, Illinois  60637, USA}
\author{Alessandro~Melchiorri}
\affiliation{Physics Department and INFN, Universita' di Roma 
	``La Sapienza'', Ple.\ Aldo Moro 2, 00185, Rome, Italy}
\noaffiliation
\author{Olga~Mena$^3$}
\noaffiliation
\author{Paolo~Serra$^2$}
\noaffiliation

\begin{abstract}

Inflation predicts primordial scalar perturbations with a nearly scale-invariant
spectrum and a spectral index \textit{approximately} unity (the
Harrison--Zel'dovich (HZ) spectrum).  The first important step for inflationary
cosmology is to check the consistency of the HZ primordial spectrum with current
observations. Recent analyses have claimed that a HZ primordial spectrum is
excluded at more than $99\%$ c.l.  Here we show that the HZ spectrum is only
marginally disfavored if one considers a more general reionization scenario. 
Data from the Planck mission will settle the issue.

\end{abstract}


\maketitle

%%%%%%%%%%%%%%%%%%%%%%%%%%%%%%%%%%%%%%%%
%%%%%%%%%%%%%%%%%%%%%%%%%%%%%%%%%%%%%%%%

%%%%%%%%%%%%%%%%%%%%%%%%%%%%%%%%%%%%%%%%
%%%%%%%%%%%%%%%%%%%%%%%%%%%%%%%%%%%%%%%%
%\section{Introduction}
%%%%%%%%%%%%%%%%%%%%%%%%%%%%%%%%%%%%%%%%
%%%%%%%%%%%%%%%%%%%%%%%%%%%%%%%%%%%%%%%%

One usually models the dynamics of primordial inflation by the dynamics of a
scalar field, know as the inflaton, evolving under the influence of a scalar
potential. The scalar and tensor perturbation spectra produced by inflation
depend upon the value of the inflaton potential during inflation and how rapidly
the scalar field evolves during inflation. Observational information of the
spectra can be translated into information about the inflaton potential, thus
about physics far beyond the scales of the standard model of particle physics
\cite{Lidsey:1995np}.   

Cosmic microwave background (CMB) observables that map onto the the potential
and its slope are 1) $n$, the spectral index of comoving scalar perturbations
$P_{\cal R}$: $n-1=d\ln P_{\cal R}/d\ln k$, and 2) $r$, the ratio of tensor
(gravitational wave) perturbations to $P_{\cal R}$ \cite{Dodelson:1997hr}. 

CMB temperature and polarization anisotropies are the most promising route to
explore the physics behind inflation.  Recent measurements by the Wilkinson
Microwave Anisotropy Probe (WMAP) satellite seven-year mission \cite{wmap7},
combined with the ground-based and balloon-borne experiments such as BOOMERanG
\cite{boom03}, QUAD \cite{quad}, ACBAR \cite{acbar}, and BICEP \cite{bicep} 
have sharpened our knowledge of some key inflationary parameters.

With regard to the dynamics of inflation, a hotly debated question is whether
the case of $n=1$ is significantly excluded by current observations (see
\textit{e.g.,} Refs.\ \cite{wmap7}-\cite{finelli}).

The scalar spectral index, $n$, has been recently constrained to the value
$n=0.963\pm0.014$ at $68\%$ c.l.\ by the WMAP seven-year dataset (WMAP7)
\cite{wmap7}, disfavoring the value of $n=1$ at about two standard deviations. A
combined analysis with galaxy clustering data gives $n=0.963\pm0.012$ at $68
\%$ c.l., ruling out $n=1$ at more than three standard deviations
 (\cite{wmap7}, \cite{finelli}).

Compelling evidence that $n \neq 1$ would be quite revealing for two reasons.
First, it would provide an indication for the dynamical evolution of the
expansion rate of the universe as perturbations are being produced. Secondly, in
many models of slow-roll inflation the amplitude of the tensor perturbations are
proportional to $|n-1|$.  Thus, the larger the departure of $n$ from unity, the
more likely tensor perturbations would be within observational reach.  

As it is well known, a scalar spectral index with an exact value of $n=1$
corresponds to the phenomenological model proposed by Harrison, Zel'dovich, and
Peebles \cite{hz}.\footnote{For the purpose of our paper, we will assume that
the tensor modes are small enough as to be irrelevant.}  While one can construct
inflationary models that give $n=1$ either approximately
\cite{Vallinotto:2003vf} or exactly \cite{Starobinsky:2005ab}, they are less
than compelling.  In fact, observations pointing to $n$ exactly unity may
indicate that the origin of cosmic perturbations lies in some unknown
fundamental process and may not arise from inflation. 

In this paper we show that the value of $n$, and its uncertainty, derived from
CMB datasets are sensitive to how one models the reionization of the universe. 
In particular, the statement that $n=1$ is observationally excluded is very much
weakened if one treats reionization in a general manner. 

It is well known from a large set of astrophysical observables that after
primordial recombination (which occurred at a redshift of $z\sim1100$) the
universe ``reionized'' at a redshift $z>6$. Current constraints from CMB data
assume a sudden and complete reionization at a redshift $z_r$. This parameter is
usually assumed to be in the range $4 < z_r < 32$, and the constraints on $n$
are obtained after  marginalization over $z_r$. The electron ionization fraction
$x_e(z)$ is parametrized by $z_r$ such that for $z \ll z_r$ $x_e(z)=1$
($x_e(z)=1.08$ for $z<3$ in order to take into account Helium recombination) and
$x_e(z)=2 \times 10^{-4}$ for $z>z_r$, \textit{i.e.,} joining the value after
primordial recombination with a smooth interpolation. 
We refer to this common procedure in what follows as ``sudden'' reionization.

However, the precise details of the reionization process are not very
well known, and therefore the reionization history of the universe at those
redshifts could have easily been very different. 

As already remarked in Refs.\ \cite{mortonson} and \cite{lewis}, the assumption
of a more general reionization scheme could affect the cosmological constraints
on $n$. It is therefore timely, especially in view of the recent CMB and galaxy
clustering data that have improved the bound on $n$, to investigate the impact
of the reionization history on the current constraints on $n$ in a more general
reionization scenario. 

We adopt two methods for parametrizing the reionization history. The first
method, developed in Ref.\ \cite{mortonson}, is based on principal components
that provide a complete basis  for describing the effects of reionization on
large-scale $E$-mode polarization.  Following Ref.\ \cite{mortonson}, one can
parametrize the reionization history  as a free function of redshift by
decomposing $x_e(z)$ into its principal components:
\begin{equation}
x_e(z)=x_e^f(z)+\sum_{\mu}m_{\mu}S_{\mu}(z),
\end{equation}
where the principal components, $S_{\mu}(z)$,  are the eigenfunctions of the
Fisher matrix that describes  the dependence of the polarization spectra on
$x_e(z)$ (again, see Ref.\ \cite{mortonson}), $m_{\mu}$ are the amplitudes of
the  principal components for a particular reionization history, and  $x_e^f(z)$
is the WMAP fiducial model at which the Fisher matrix is computed and from which
the principal components are obtained. In what follows we use the publicly
available $S_{\mu}(z)$ functions and varied the amplitudes $m_{\mu}$ for
$\mu=1,...,5$ for the first five eigenfunctions. Hereafter we refer to this
method as the MH (Mortonson-Hu) case.

In the second approach, we use a different parametrization, sampling the
evolution of the ionization fraction $x_e$ as a function of the redshift $z$ at
$7$ points ($z=9,12,15,18,21,24,27$), and interpolating the value of $x_e(z)$
between them with a cubic spline. For  $30 < z$ we fix $x_{e}=2\times
10^{-4}$ as the value of $x_{e}$ expected before reionization (and after
primordial recombination),  while $x_e = 1$ for $3<z<6$ and $x_e=1.08$ for $z <
3$  in order to be in agreement  with both Helium ionization and Gunn-Peterson
test observations. This approach is very similar to the one used in Ref.\
\cite{lewis}, and we will refer to it as the LWB (Lewis-Weller-Battye) case.

Exploiting two different and more general reionization schemes will provide us
with not only a consistency check of the results, but also a broader set of
reionization histories. Finally, we also consider here the standard reionization
model, \textit{i.e.,} sudden reionization.

We have then modified the Boltzmann CAMB code \cite{camb} incorporating the two
generalized reionization scenarios and extracted cosmological parameters from
current data using a Monte Carlo Markov Chain (MCMC) analysis based on the
publicly available MCMC package \texttt{cosmomc} \cite{Lewis:2002ah}.

We consider here a flat $\Lambda$CDM universe described by a set of cosmological
parameters
\begin{equation}
 \label{parameter}
      \{\omega_b,\omega_c,
      \Theta_s, n, \log[10^{10}A_{s}] \},
\end{equation}
where $\omega_b\equiv\Omega_bh^{2}$ and $\omega_c\equiv\Omega_ch^{2}$ are the
physical baryon and cold dark matter densities  relative to the critical
density, $\Theta_{s}$ is the ratio of the sound horizon to the angular diameter
distance at decoupling, $A_{s}$ is the amplitude of the primordial spectrum, and
$n$ is the scalar spectral index. In one case we will also consider the possibility 
of a redshift-dependent dark energy  component evolving with equation of state
$w(z)=w_0+w_z z/(1+z)$.

The extra parameters needed to describe the reionization are the five amplitudes
of the eigenfunctions for the MH case, or the five amplitudes in the five bins
for the LWB case, and one single common parameter, the optical depth, $\tau$,
for the sudden reionization case. 

Our basic data set is the seven--year WMAP data \cite{wmap7}  (temperature and
polarization) with the routine for computing the likelihood supplied by the WMAP
team. Together with the WMAP data, we also augment the WMAP7 data with the CMB
datasets from BOOMERanG \cite{boom03}, QUAD \cite{quad}, ACBAR \cite{acbar}, and
BICEP \cite{bicep}. For all these experiments we marginalize over a possible
contamination from the Sunyaev-Zeldovich component, rescaling the WMAP template
at the corresponding experimental frequencies. We therefore consider two cases:
we first analyze the WMAP data alone,  referring to it as to the  ``WMAP7''
case, and we then include the remaining CMB experiments (``CMB All''). Finally
we also consider, separately, the galaxy clustering results from LRG-7 of Ref.\
\cite{reid} and  the baryonic acoustic oscillation (BAO) data from the same
dataset, see Ref.\ \cite{percival}.

%%%%%%%%%%%%%%%%%%%%%%%%%%%%%%%%%%%%%%%%
\begin{figure}[!ht]
\includegraphics[width=7.2cm]{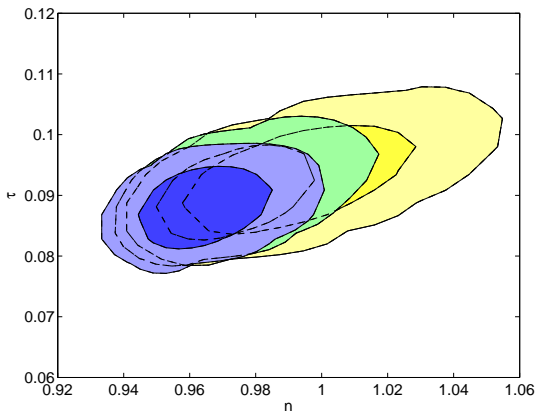}

\caption{\label{fig1}Contours of the $68 \%$ and $95 \%$ c.l.\ constraints in
the $n$ vs $\tau$ plane for different datasets. The contours regions come from a
generalized reionization scenario using (from right to left) the MH approach 
from WMAP-7 data (Yellow), ``CMB all'' (Green), All CMB+BAO (Blue).}

\end{figure}
%%%%%%%%%%%%%%%%%%%%%%%%%%%%%%%%%%%%%%%%

Our main results are reported in Table I, where we list the constraints on $n$
for the different reionization scenarios using different datasets.  Notice that
using the MH parametrization drastically alters the constraints on $n$. For the
WMAP7 case alone, the HZ spectrum  is not only in agreement with the data, but
is also close to the best-fit value.  When the remaining CMB experiments are
included, the best-fit value of $n$ shifts to lower values, but it is still 
consistent at better than 95\% c.l.\ with $n=1$. When information from galaxy
clustering is included, a value of $n=1$ is excluded at about the $95\%$ c.l.,
\textit{i.e.,} in a much less conclusive way than in the sudden case
where $n=1$ is excluded at more than $99.7\%$ c.l.\ (see \textit{e.g.,} Ref.\
\cite{finelli}).

Adopting the LWB parameterization of reionization has very similar effects. The
HZ spectrum is only marginally excluded at a 95\% confidence level if the
complete CMB dataset is considered. Notice that in the case of the LWB
parametrization, the results are closer to those obtained in the standard
sudden reionization scheme. Therefore, the LWB parameterization is less
general and samples fewer reionization scenarios than the MH method.

%%%%%%%%%%%%%%%%%%%%%%%%%%%%%%%%%%%%%%%%
\begin{table}[!ht]
\begin{ruledtabular}
\begin{tabular}{lcrr}
Dataset & Ionization & $n$ ($68 \%$ c.l.) & $95\%$ c.l. \\
\hline
\hline
WMAP7   & sudden & $0.965\pm 0.014$ & $n\le 0.993$ \\
\vspace{0.1cm}
CMB All & sudden &  $0.959\pm 0.013$ & $n\le 0.984$  \\
\vspace{0.1cm}
WMAP7  & MH & $0.993\pm0.023$ & $n\le 1.042$ \\
\vspace{0.1cm}
CMB All & MH & $0.975\pm0.017$ & $n\le 1.011$ \\
\vspace{0.1cm}
CMB All+LRG-7 & MH & $0.966\pm0.014$ & $n\le 0.994$ \\
\vspace{0.1cm}
CMB All+BAO & MH & $0.965\pm0.014$ & $n\le 0.995$ \\
\vspace{0.1cm}
CMB All+BAO & MH+$w(z)$ & $0.985\pm0.018$ & $n\le 1.025$ \\
\vspace{0.1cm}
WMAP7  & LWB & $0.977\pm0.018$ & $n\le 1.01\phantom{0}$ \\
\vspace{0.1cm}
CMB All & LWB &  $0.963\pm0.015$ & $n\le 0.998$ \\
\end{tabular}
\end{ruledtabular}

\caption{Constraints and upper limits on $n$ from several datasets in different recombination
scenarios.}

\label{table1}
\end{table}
%%%%%%%%%%%%%%%%%%%%%%%%%%%%%%%%%%%%%%%%

Considering the WMAP7 data alone, the HZ spectrum is at $\Delta(-2\ln({\cal
L}))= 0.72$ with respect to the best fit model in the case of a MH recombination
scenario ($\Delta(-2\ln({\cal L}))=2.2$ when all CMB experiments are
considered), compared with $\Delta(-2\ln({\cal L}))= 3.3$ when recombination is
sudden ($\Delta(-2\ln({\cal L}))= 5.8$ in case of CMB All). 

It is interesting to consider the constraints on the optical depth $\tau$,
derived by integrating $x_e(z)$ up to $z=32$. Figure 1 shows the 68\% and 95\%
c.l.\ constraints in the $n$ vs.\ $\tau$ plane arising from different datasets
and assuming the MH reionization parametrization. As we can see the optical
depth is always in the range $0.08$-$0.11$, slightly higher than in the standard
analysis but consistent with  several physical reionization models. Notice also
that the HZ spectrum ($n=1$) is perfectly compatible with WMAP7. Including more
datasets shrinks and shifts the contours towards lower values of $n$, but the HZ
case is always reasonably consistent at the 95\% c.l..  Including BAO data rules
out the HZ spectrum at about $2$ standard deviations that is considerably less
stringent than in the standard case. Moreover, considering a redshift-dependent
equation of state weakens the geometrical probes and HZ is again inside the $68
\%$ confidence level for the CMB+BAO case.

It is important to investigate if a more general reionization scenario with
the assumption of $n=1$ could provide a viable cosmology, \textit{i.e.,} if the
value  of cosmological parameters such as the baryon density or the age of the
universe are compatible with complementary cosmological  information from Big
Bang Nucleosythesis and age constraints. It has been shown, for example, in
Ref.\ \cite{finelli} that the assumption of an HZ spectrum in the standard
analysis with sudden reionization gives a baryon abundance that is at
odds with current bounds from BBN. We have therefore performed an analysis using
WMAP7 and assuming $n=1$ for the MH reionization scenario, and the results for
the different cosmological parameters are summarized in Table 2.

%%%%%%%%%%%%%%%%%%%%%%%%%%%%%%%%%%%%%%%%
\begin{table}[!ht]
\begin{ruledtabular}
\begin{tabular}{lr}
Parameter & Constraint ($68\%$ c.l.)\\
\hline \hline
\vspace{0.1cm}
$\Omega_bh^2$ &  $ 0.0234\pm 0.0004$\\
\vspace{0.1cm}
$\Omega_ch^2$& $ 0.106\pm 0.005$\phantom{0}\\
\vspace{0.1cm}
$H_0$ (km s$^{-1}$Mpc$^{-1}$) & $ 74.2\pm 2.1$\phantom{000}\\
\vspace{0.1cm}
Age (Gyr) & $13.6\pm 0.1$\phantom{000}\\
\vspace{0.1cm}
$\Omega_{\Lambda}$ &  $0.765\pm 0.022$\phantom{0}\\
\end{tabular}
\end{ruledtabular}

\caption{Constraints on cosmological parameters from WMAP7 assuming a HZ primordial 
spectrum and a reionization scenario characterized by the MH parametrization.}

\label{table2}
\end{table}
%%%%%%%%%%%%%%%%%%%%%%%%%%%%%%%%%%%%%%%%

Notice from the results of Table $2$ that the constraints on the remaining 
cosmological parameters are all viable when compared with independent 
cosmological observables. Current Big Bang Nucleosynthesis constraints  from
observations of Deuterium abundances provide a constraint of
$\Omega_bh^2=0.0213\pm 0.002$ at $95 \%$ c.l.\ \cite{bbn}, which is compatible
with the WMAP7 constraint assuming HZ. The constraints shown in Table $2$ on
$H_0$ are also well in agreement with the recent HST determination $h=0.748\pm
0.036$ at $68 \%$ c.l.\ \cite{riess}. Figure 2 depicts the constraints from
WMAP7 in the $\Omega_bh^2$ vs.\ $H_0$ plane in a generalized recombination
scenario, assuming both a varying $n$ and a fixed $n=1$ scenarios.  Notice that
fixing $n=1$ implies larger values of both $\Omega_bh^2$ and $H_0$.

%%%%%%%%%%%%%%%%%%%%%%%%%%%%%%%%%%%%%%%%
\begin{figure}[!ht]
\includegraphics[width=7.2cm]{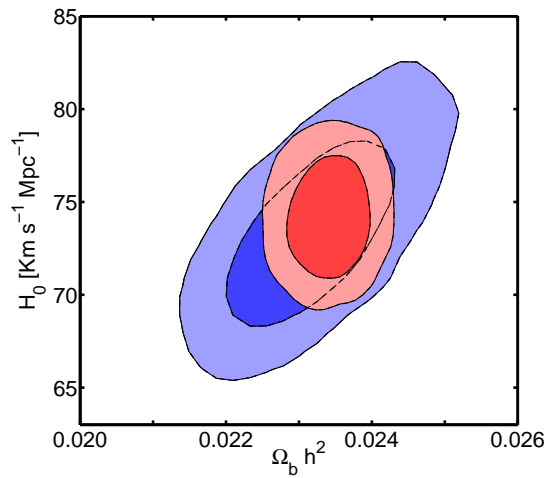}

\caption{\label{fig2}$68 \%$ and $95 \%$ c.l.\ WMAP7 constraints in the $H_0$
vs.\ $\Omega_bh^2$ plane in a generalized reionization scenario described by the
MH parametrization. The larger (blue) contours regions arise when $n$ is
allowed to vary. The smaller (red) contours correspond to the HZ ($n=1$)
spectrum.}

\end{figure}
%%%%%%%%%%%%%%%%%%%%%%%%%%%%%%%%%%%%%%%%

In conclusion, the details of the reionization processes in the late universe
are not very well known. In the absence of a precise, full redshift evolution
description of the ionization fraction during the reionization period, a simple
parametrization, with a single parameter $z_r$, has become the standard
reionization scheme in numerical analyses. However, more general reionization
scenarios are certainly plausible and their impact on the cosmological
constraints should be carefully explored. For example, processes of dark matter
annihilation as in Ref.\ \cite{galli} could easily be present. 

In this paper we have investigated the stability of the CMB constraints on $n$ 
in generalized reionization scenarios. Our study is motivated by the recent
claim that an HZ spectrum is ruled out by current WMAP data at more than $95\%$
c.l. In agreement with previous studies, we have found that a more general
treatment of reionization drastically weakens current CMB constraints on $n$. If
only the WMAP 7-year data is considered, a HZ $n=1$ spectrum lies well within
the 68\% c.l.\ allowed region.  If all current CMB datasets are considered in
the analysis, the HZ spectrum still lies well within the 95\% c.l.\ allowed
region. If the assumption of constant dark energy is relaxed, HZ is again inside
the 68\% c.l.  Therefore, current data do not yet rule out an HZ spectrum in a
conclusive way. It is also worth mentioning that non standard processes during 
recombination (see \textit{e.g.,} Ref.\ \cite{galli2}) or extra relativistic
particles (see \textit{e.g.,} Ref. \cite{bowen})  could be present and could
also put $n=1$ in agreement with current observations. 

Near future data from the Planck satellite mission are clearly needed to solve
this important question. In this respect, performing a parameter analysis  with
simulated data with the experimental configuration described in Ref.\
\cite{bluebook}, and assuming a model with $n=0.965$, we found that Planck will 
be able to discriminate it from a HZ spectrum at more than five standard
deviations even in the generalized reionization cases describes here.

%%%%%%%%%%%%%%%%%%%%%%%%%%%%%%%%%%%%%%%%
%%%%%%%%%%%%%%%%%%%%%%%%%%%%%%%%%%%%%%%%

%%%%%%%%%%%%%%%%%%%%%%%%%%%%%%%%%%%%%%%%
%%%%%%%%%%%%%%%%%%%%%%%%%%%%%%%%%%%%%%%%

\end{document}